\documentclass{elsart4}
\usepackage{graphicx}
\usepackage{dcolumn}
\usepackage{bm}
\usepackage{amsmath}
\usepackage{amsfonts}
\usepackage{amssymb}

\begin{document}

\begin{frontmatter}

\title{Effect of carbon nanotube on ballistic conduction through single-quantum-dot}

\author[IFRJ]{Thiago Lobo},
\ead{thiago.fonseca@ifrj.edu.br}
\author[UFAM]{Minos A. Neto},
\author[IFAM]{Marcio G. da Silva},
\author[UFAM]{Octavio D. R. Salmon},

\address[IFRJ]
{Instituto Federal de Educa\c{c}\~ao, Ci\^{e}ncia e Tecnologia do Rio de Janeiro - IFRJ\\
Rua Dr. Jos\'e Augusto Pereira dos Santos, s/n (C.I.E.P. 436 Neusa Brizola), 
24425-004 S\~ ao Gon\c{c}alo, Rio de Janeiro, Brasil.} 

\address[UFAM]
{Departamento de F\'{\i}sica, Universidade Federal do Amazonas, 3000, Japiim,
69077-000, Manaus-AM, Brazil}

\address[IFAM]
{Instituto Federal do Amazonas, Av. 7 de Setembro, 1975, Centro, Manaus-AM, Brazil}


\begin{abstract}

We will study the competitive effect between the transport of a quantum dot adsorbed to a ballistic channel and laterally coupled to a single-walled carbon nanotube (SWNT).
We will use the ``tight-binding" approach to analytically write the SWNT Green's function and the quantum dot will be solved by the atomic method for $U\rightarrow\infty$.
We will present curves of the electronic density of states for some different sizes of nanotubes. The results for the conductance curves will be presented as a function of $E_ {f}$ and for different values of $n$ and hopping between the nanotube and the quantum dot .

\end{abstract}


\end{frontmatter}

\section{Introduction}
\label{Sec1}

In the last few decades, low-dimension systems have been the focus of great interest on the part of the scientific community. Due to their extraordinary physical properties, these carbon-based systems have applications in areas ranging from medicine: such as DNA nanosensors \cite{nanomat2006}, to nanoelectronics: with quantum dot systems interacting with graphene \cite{chen2015} and carbon nanotubes \cite{thiago2020}.

Quantum mechanics combined with epitaxial molecular growth and chemical deposition techniques are of fundamental importance in understanding and making quantum dots. An example of this device is the tunnel diode \cite{esaki1992}, which can play an important role in the development of quantum dot circuits \cite{childress2004,delbecq2011,frey2012}.

From the theoretical point of view \cite{meir1993}, these low-dimensional systems have revealed several fundamental aspects in the understanding of the interaction between light and matter, such as the development of circuits in Quantum Electrodynamics - \textit {QED Circuits}\cite{wallraff2004}. In addition, quantum dots have electronic properties at low temperatures governed by the Kondo effect, which can be observed both in the study of theoretical models \cite{aligia2002} as well as in experimental results \cite{gordon1998,cronenwet1998}.

Recently, quantum junctions - a combination of a quantum dot coupled to a carbon nanotube - have demonstrated a probe for transferring charges between discrete and continuous \cite{bruhat2016}. This type of coupling was presented in systems that can be used in high conductance devices, laterally connecting a quantum dot with a sheet of graphene \cite{chen2015}.

In this work, we present a device that combines with a quantum dot adsorbed in a ballistic channel, coupled laterally to a carbon nanotube. We are interested in the properties of transport through the ballistic channel, which at low temperatures is governed by the Kondo effect at the quantum dot. The carbon nanotube, connected to the quantum dot through a hopping energy, also affects these transport properties, since it injects conduction electrons into the system that produce a destructive effect at the Kondo peak.

\begin{figure}[htbp]
\centering
\includegraphics[width=7.5cm,height=6.0cm,angle=0.0]{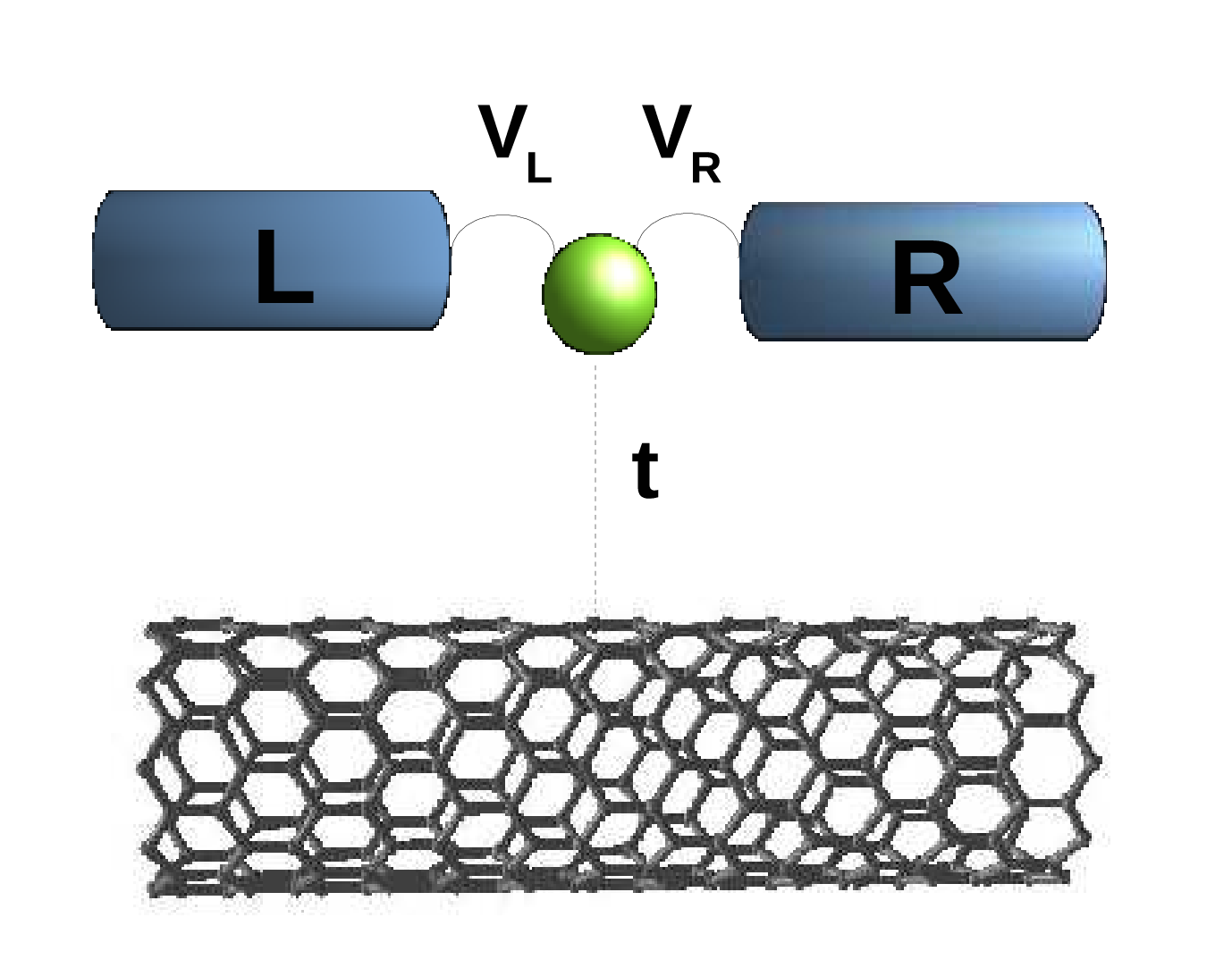}
\caption{Quantum dot adsorbed in a ballistic channel laterally coupled to a carbon nanotube.} 
\label{fig1}
\end{figure}

We apply the atomic approach \cite{Nanotech1} to model the quantum dot and use the Dyson's equation to connect the quantum dot to the ballistic channel and carbon nanotube, as schematically represented in Fig.(\ref{fig1}).

In Section $2$ we present the model and the formalism of Green's function (GF). The results and discussions are described in Section $3$. As a final point, the last section is devoted to final remarks and conclusions.

\section{Model and Formalism}
\label{sec2} 

The quantum dot was modeled by the atomic approach \cite{Thiago1_2006qd,Nanotech1} and the leads at the points ($L$ and $R$) were represented by two square bands where the Green Function can written as
\begin{equation}
G^{0}_{c}(\textbf{k}, z)=\frac{-1}{z-E(\textbf{k})}
\label{Eq1}
\end{equation}
where $E(\textbf{k})$ is the energy of the free electrons with wave vector $\textbf{k}$, which must be integrated for all $\textbf{k}$ values, in the range from $-D$ to $D$, where $D$ is the half-width of the conduction band.

The GF of the zigzag nanotube is calculated analytically, 
adopting a single $\pi-$band tight-binding Hamiltonian approach \cite{Thiago2_2006nt}.

\begin{equation}
G^{SWNT}_{jj'}(z)=\frac{1}{n}\sum_{l}\frac{a\sqrt{3}}{2\pi}\int^{\pi/a\sqrt{3}}_{-\pi/a\sqrt{3}}
dk_{x}\frac{z.e^{i.\stackrel{\rightarrow}{k}.(\stackrel{\rightarrow}{R_{j}}-\stackrel{\rightarrow}
{R_{j'}})}}{z^{2}-\epsilon^{+}(\tilde{k_{x}})}
\label{Eq2}
\end{equation}
where the energy dispersion relation $\epsilon^{\pm}(\tilde{k}_{x})$ for the zigzag nanotube, may be written as\cite{livro}

\begin{equation}
\epsilon^{\pm}(\tilde{k}_{x})=\pm t \sqrt{1+4\cos\left(\tilde{k}_{y}\right)\cos(\tilde{k}_{x})+4\cos^{2}\left(\tilde{k}_{y}\right)}. 
\label{Eq3}
\end{equation}%
with $\tilde{k}_{x}=k_{x}a\sqrt{3}/2$, $\tilde{k}_{y}={k}_{y}a/2$ and for $-\pi/\sqrt{3} < k_{x}a < \pi/\sqrt{3}$.
For the zigzag nanotube $(n,0)$, the ${k}_{y}=\frac{2\pi l}{an}$ is quantized, where $l=1,...,n$, $a$ is the lattice parameter, whose approximate value is of $2.46$ \AA{} and $t$ is the electronic hopping between the carbon atoms, whose value we consider equal to the unit.

The density of states (DOS) from the nanotube's conduction band is calculated using
\begin{equation}
\rho_{c}(z)=-\frac{1}{\pi}\textit{Im}\left(G^{SWNT}_{jj}(z)\right)
\label{Eq4}
\end{equation}

The quantum dot adsorbed in the ballistic channel, in the absence of the carbon nanotube, can be modeled by the Anderson impurity model with infinite Coulomb repulsion $(U\rightarrow\infty)$. The coupling between the nanotube and the quantum dot will be carried out later with the Dyson's equation method. The Hamiltonian is given by

\begin{eqnarray}
\mathcal{H} &=&\sum_{\mathbf{\alpha k},\sigma }E_{\mathbf{\alpha k},\sigma }c_{\mathbf{\alpha k},\sigma}^{\dagger }c_{\mathbf{\alpha k},\sigma }+\sum_{\sigma }\ E_{f,\sigma }X_{f,\sigma\sigma } \nonumber \\
&&+\sum_{\mathbf{\alpha k},\sigma}V_{\mathbf{\alpha k}}\left(X_{f,0\sigma}^{\dagger }c_{\mathbf{\alpha k},\sigma }+H.c.\right),
\label{Eq5}
\end{eqnarray}
where the first term represents the conduction electrons ($c$-electrons) from the left ($\alpha = L$) and right ($\alpha = R$) electrodes , the second describes the quantum dot characterized 
by a localized $f$ level $E_{f,\sigma}$ and the last one corresponds to the interaction between the $c$-electrons and the quantum dot. For the sake of simplification, we consider a constant hybridization $V_{\mathbf{\alpha k}}= V$. 

For the case with infinite Coulomb repulsion $(U\rightarrow\infty)$, Hubbard operators \cite{FFM} are employed to project out the double occupation state $\left|f,2\right\rangle$ from the local states at the quantum dot. Local projectors 
into the states $|f,a\rangle$ at the quantum dot, must respect the following identity:
$X_{f,00}+X_{f,\sigma\sigma}
+X_{f,\overline{\sigma}\overline{\sigma}}=I$, where $\overline{\sigma}=-\sigma$. Being that $a=0$, $\sigma$ or $\overline{\sigma}$, the occupation numbers on the quantum dot 
$n_{f,a}=\langle X_{f,aa}\rangle$ should then satisfy the ``completeness'' relation

\begin{equation}
n_{f,0}+n_{f,\sigma}+n_{f,\overline{\sigma}}=1.  
\label{Eq6}
\end{equation}%

\subsection{The quantum dot}

The quantum dot's GF ($G_{qd,\sigma}$) presented in this study is based on the Hubbard cumulative expansion technique, using hybridization as a perturbation \cite{FFM,FFF}. Considering $M_{\sigma}^{eff}$ as an effective cumulant that cannot be separated by cutting a single edge from Feynmann diagram (irreducible diagram) and $G^{0,\alpha}_{c}(\mathbf{k}{,}z))$ as the GF of the conduction electrons of electrodes ($\alpha$= $L$,$R$), We can write the exact GF for the  $ f $ electron located at the quantum dot as follows

\begin{equation}
G_{qd,\sigma}=M_{\sigma}^{eff}+M_{\sigma}^{eff} V G^{0,\alpha}_{c} V M_{\sigma}^{eff}
\nonumber
\end{equation}
\begin{equation}
+M_{\sigma}^{eff} V G^{0,\alpha}_{c} V M_{\sigma}^{eff} V G^{0,\alpha}_{c} V M_{\sigma}^{eff}+....
\label{Eq7}
\end{equation}

\begin{equation}
G_{qd,\sigma}=M_{\sigma}^{eff}\left[1+ G^{0,\alpha}_{c}V^{2} M_{\sigma}^{eff}
+\left(G^{0,\alpha}_{c} V^{2} M_{\sigma}^{eff}\right)^{2}+...\right].
\label{Eq8}%
\end{equation}

The term in square brackets represents the infinite sum of a geometric progression, with a ratio equal to $G^{0,\alpha}_{c}V^{2} M_{\sigma}^{eff}$. The final result is given by

\begin{equation}
G_{qd,\sigma}(z)=\ \frac{M_{\sigma}^{eff}(z)}{1-M_{\sigma}^{eff}(z)V^{2}\sum_{\alpha\mathbf{k}}G^{0,\alpha}_{c}(\mathbf{k},z)\ }.
\label{Eq9}%
\end{equation}
where the sum over the two identical channels $\alpha$ results in multiplier $2$.

The calculation of the exact value of $M_{\sigma}^{eff}(z)$ is an impracticable task, however, we will replace it by 
an atomic cumulant $M_{\sigma}^{at}(z)$, obtained from the exact solution of the Anderson's impurity in the atomic limit, 
when the band of uncorrelated electrons has zero width (atomic model).

The atomic solution will be obtained in the next section. We call the replacement of $M_{\sigma}^{eff}(z)$ by $M_{\sigma}^{at}(z)$ an ``atomic approach''.

\subsection{The atomic solution}

The atomic solution of Anderson's impurity problem at the limit of infinite Coulomb repulsion $U\rightarrow\infty$, can be obtained by diagonalizing the Hamiltonian Eq.(\ref{Eq5}) by the states $\left|f,c\right\rangle$, where the possibilities are: $\left|f\right\rangle=0, \uparrow, \downarrow$ and 
$\left|c\right\rangle=0, \uparrow, \downarrow, \uparrow\downarrow$. The GF of the atomic model is obtained by the Zubarev equation \cite{Zubarev}

\begin{eqnarray}
G^{at}(z) &=& e^{\beta \Omega}\sum_{n}\sum_{jj^{\prime}}e^{-\beta E_{n,j}}+e^{-\beta E_{n-1,j^{\prime}}} \nonumber \\
&&\times\frac{\left|\left\langle n-1,j^{\prime}\left|X_{\mu}\right|n,j\right\rangle\right|^{2}}{z-(E_{n,j}-E_{n-1,j^{\prime}})},
\label{Eq10}
\end{eqnarray}

where $\Omega$ is the thermodynamical potential and the eigenvalues $E_{n,j}$ and eigenvectors $|nj\rangle$ correspond to 
the complete solution of the Hamiltonian Eq.(\ref{Eq5}). The different transitions occur between states with $n$ and 
$n + 1$ particles that satisfy $\langle n-1, j^{\prime}|X_{\mu}|n,j\rangle\neq 0$. The final result is the following:

\begin{equation}
G^{at}(z)=e^{\beta \Omega}\sum^{8}_{i=1}\frac{m_{i}}{z-u_{i}}.
\label{Eq11}
\end{equation}

The complete calculation of all poles $u_{1}$ and residues $m_{i}$ is detailed in reference \cite{Nanotech1}.

\subsection{The atomic approach} 

On the other hand, following the same procedure described for obtaining $G_{qd,\sigma}(z)$, the exact atomic $f$ GF has the same 
form of Eq.(\ref{Eq9}) and can be written as \cite{FFF}: 

\begin{equation}
G_{\sigma}^{at}=M_{\sigma}^{at}+M_{\sigma}^{at} V G^{o}_{c} V M_{\sigma}^{at}
\nonumber
\end{equation}
\begin{equation}
+M_{\sigma}^{at} V G^{o}_{c} V M_{\sigma}^{at} V G^{o}_{c} V M_{\sigma}^{at}+....
\label{Eq12}%
\end{equation}

\begin{equation}
G_{\sigma}^{at}=M_{\sigma}^{at}\left[1+ G^{o}_{c}V^{2} M_{\sigma}^{at}
+\left(G^{o}_{c} V^{2} M_{\sigma}^{at}\right)^{2}+...\right].
\label{Eq13}
\end{equation}

\begin{equation}
G_{\sigma}^{at}(z)=\frac{M_{2,\sigma}^{at}(z)}{1-M_{2,\sigma}^{at}(z)\mid V\mid^{2}\sum_{\mathbf{k}}G_{c,\sigma}^{o\bar{e}}(\mathbf{k},z)},\label{Eq14}
\end{equation}

where $G_{c,\sigma}^{o\bar{e}}(\mathbf{k},z)=-1/(z-\varepsilon(\mathbf{k}))$. From this equation we then obtain an explicit expression 
for $M_{2,\sigma}^{at}(z)$ in terms of $G_{ff,\sigma}^{at}(z)$ 

\begin{equation}
M_{\sigma}^{at}(z)=\frac{G_{\sigma}^{at}(z)}{1+G_{\sigma}^{at}(z)V^{2}\sum_{\mathbf{k}}G_{c,\sigma}^{o}(\mathbf{k},z)},
\label{Eq15}
\end{equation}%
where $G_{\sigma}^{at}$ was calculated analytically and is given by Eq.(\ref{Eq11}). 

The atomic approach consists in substituting $M_{\sigma}^{eff}(z)$ in Eq.(\ref{Eq9}) by the $M_{\sigma }^{at}(z)$, 
given by Eq.(\ref{Eq15}). Finally, the approximate GF of the quantum dot connected to leads $L$ and $R$ is written as

\begin{equation}
G_{qd,\sigma}(z)= \frac{M_{\sigma}^{at}(z)}{1-M_{\sigma}^{at}(z)\Delta^{2}\sum_{\alpha,\mathbf{k}}G_{c,\sigma}^{o}(\mathbf{k},z)},
\label{Eq16}%
\end{equation}
where $G_{c{,}\sigma}^{o}(\mathbf{k}{,}z)$ is the GF of the leads and must be integrated in the range from $-D$ to $D$,
which is the width of the conduction band. With the atomic approach, the different energies $E_{k,\sigma}$ of the c-electrons, which should contribute to the propagators of the effective cumulant $M_{\sigma }^{eff}(z)$ , are now replaced by the contribution of a single $E_{c}$ energy 
value in $M_{\sigma }^{at}(z)$. Thus, it is essential to replace the constant  $V^{2}$ by $\Delta^{2}$, decreasing the contribution of the $c$-electrons with was overestimated by concentrating them at a single energy level.  We use $\Delta=\pi V^{2}/2D$ like the Kondo peak width, where $D$ is the half conduction bandwidth.

\subsection{Connection with the nanotube}

The system formed by the quantum dot and the leads (site $0$) must still be connected with the nanotube (site $1$). We will use the Dyson's equation method to make this. Let's use hybridization operator $\hat{V}=|0\rangle $$\bf{t}$$ \langle 1|+|1\rangle $$\bf{t}$$\langle 0|$, 
where $ \bf{t} $ is the hopping between the carbon nanotube and the quantum dot and will be given in units of $V$, from the Hamiltonian of Eq.(\ref{Eq5}).
Dressed GF at site of quantum dot $G_{00, \sigma}$ can be written in terms of the localized undressed GF $G_{jj}^{SWNT}$ of the carbon nanotube Eq.(\ref{Eq2}), and in terms of the undressed GF $G_{qd, \sigma}$ of the quantum dot with the leads Eq.(\ref{Eq16}).

\begin{equation}
\hat{G}=\hat{g}+\hat{g}\hat{V}\hat{G},
\nonumber
\end{equation}%
\begin{equation}
G_{00, \sigma}=g_{00, \sigma}+g_{00, \sigma}\textbf{t}G_{10, \sigma}+g_{01, \sigma}\textbf{t}G_{00, \sigma},
\nonumber
\end{equation}
\begin{equation}
G_{10}=g_{10, \sigma}+g_{10, \sigma}\textbf{t}G_{11, \sigma}+g_{11, \sigma}\textbf{t}G_{00, \sigma}.
\end{equation}

After solving this system of equations, and considering $g_{10, \sigma}=g_{01, \sigma}=0$, we can write

\begin{equation}
G_{00, \sigma}=\frac{g_{00, \sigma}}{(1-g_{00, \sigma}\textbf{t}^{2}g_{11, \sigma})},
\end{equation}

and finally

\begin{equation}
G_{00, \sigma}=\frac{G_{qd, \sigma}}{(1-G_{qd, \sigma}\textbf{t}^{2}G_{jj}^{SWNT})}.
\label{Eq17}
\end{equation}

\section{Results and Discussions}
\label{sec4}

\begin{figure}[h!]
	\centering
	\includegraphics[width=7.5cm,height=6.0cm,angle=0.0]{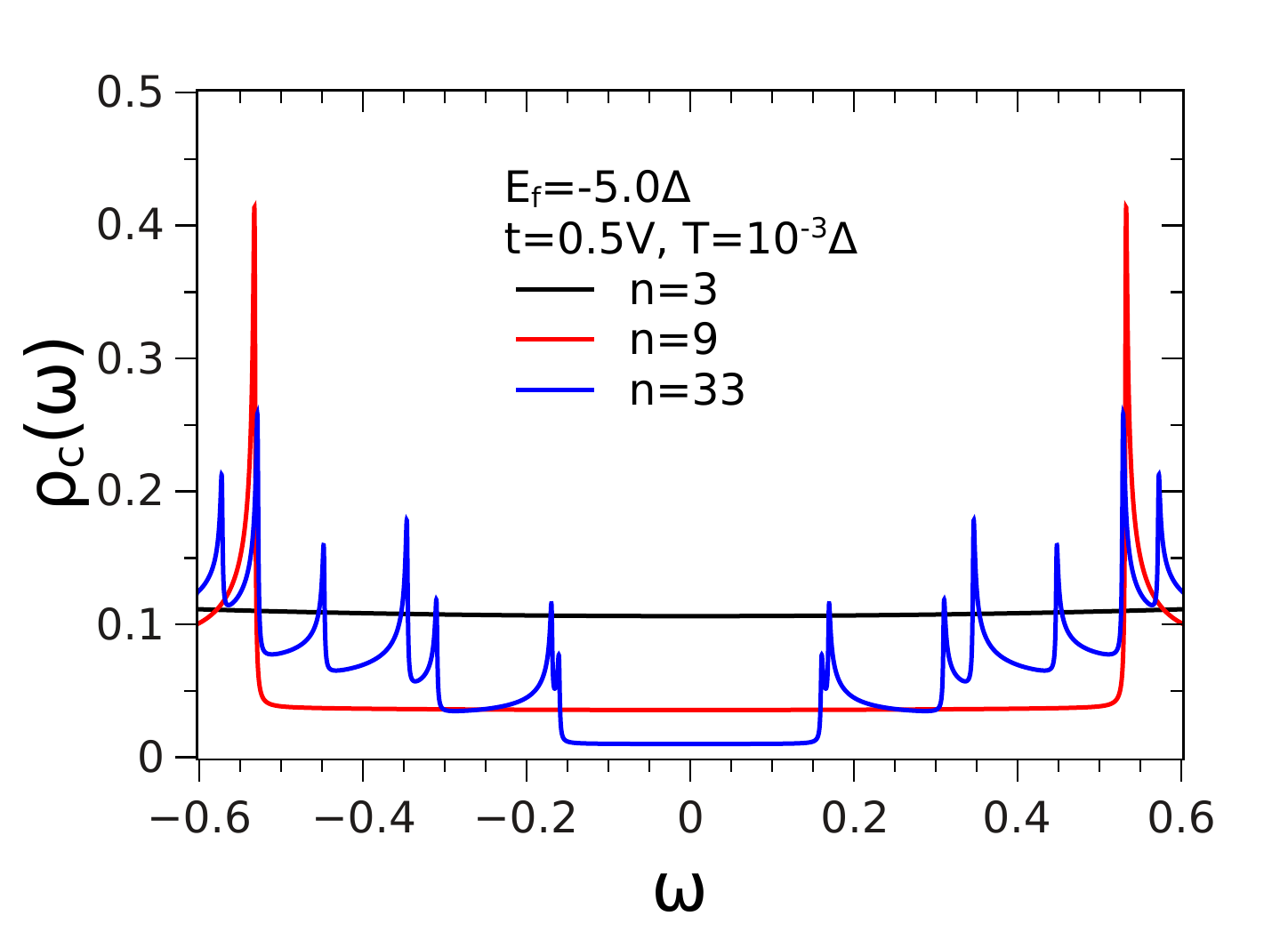}
	\caption{The density of states of the conduction electrons $\rho_{c}$  as a function of the $\omega$ (multiplied by $\Delta=0.01$) and for $n=3$, $n=9$ e $n=33$.} 
	\label{fig2}
\end{figure}

\begin{figure}[h!]
	\centering
	\includegraphics[width=7.5cm,height=6.0cm,angle=0.0]{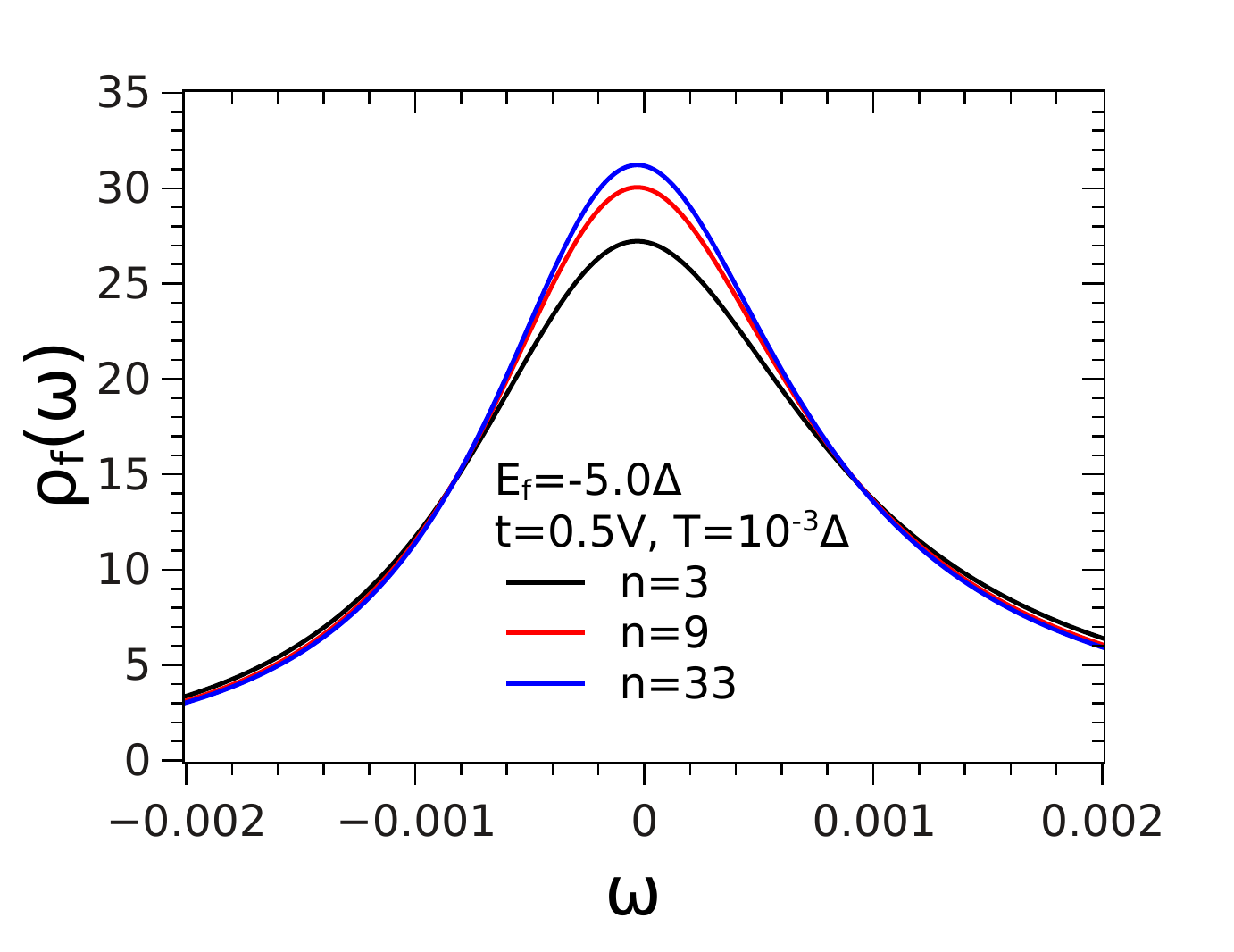}
	\caption{The density of states of the localized
		f level $\rho_{f}$ as a function of the $\omega$ (multiplied by $\Delta=0.01$) and for $n=3$, $n=9$ e $n=33$. $E_{f}=-5.0\Delta$, $\textbf{t}=0.5V$ and $T=0.001\Delta$. Detail of the height of kondo peak.} 
	\label{fig3}
\end{figure}

In the set of Fig.(\ref{fig2})-Fig.(\ref{fig3}) we will present the density of states curves of the conducting electrons of the nanotube, for different values of $n$, as well as its importance in the DOS of the electrons localized at the quantum dot, when the hopping between the SWNT and the system formed by QD $+$ balistic channel is $\textbf{t}=0.5V$.

In Fig.(\ref{fig2}) we show the density of states of the conducting electrons of the carbon nanotube for different values of $n$. The carbon nanotube of the zig zag type will be metallic for values multiple of $3$. We can see in the figure that as the size of the tube increases, represented by the increase in $n$, the height of the DOS at the Fermi level ($ E_ {F} = 0 $) decreases.

In Fig.(\ref{fig3}) we show the density of states of the localized electrons of the quantum dot (QD) for different values of $n$. The figure shows the detail of the Kondo peak, located on the chemical potential $\mu = 0$. We verified that the height of the Kondo peak increases with the increase of $n$, thus showing a behavior that is the opposite of that presented by the density of conduction electron states at the Fermi level.

\begin{figure}[h!]
	\centering
	\includegraphics[width=7.5cm,height=6.0cm,angle=0.0]{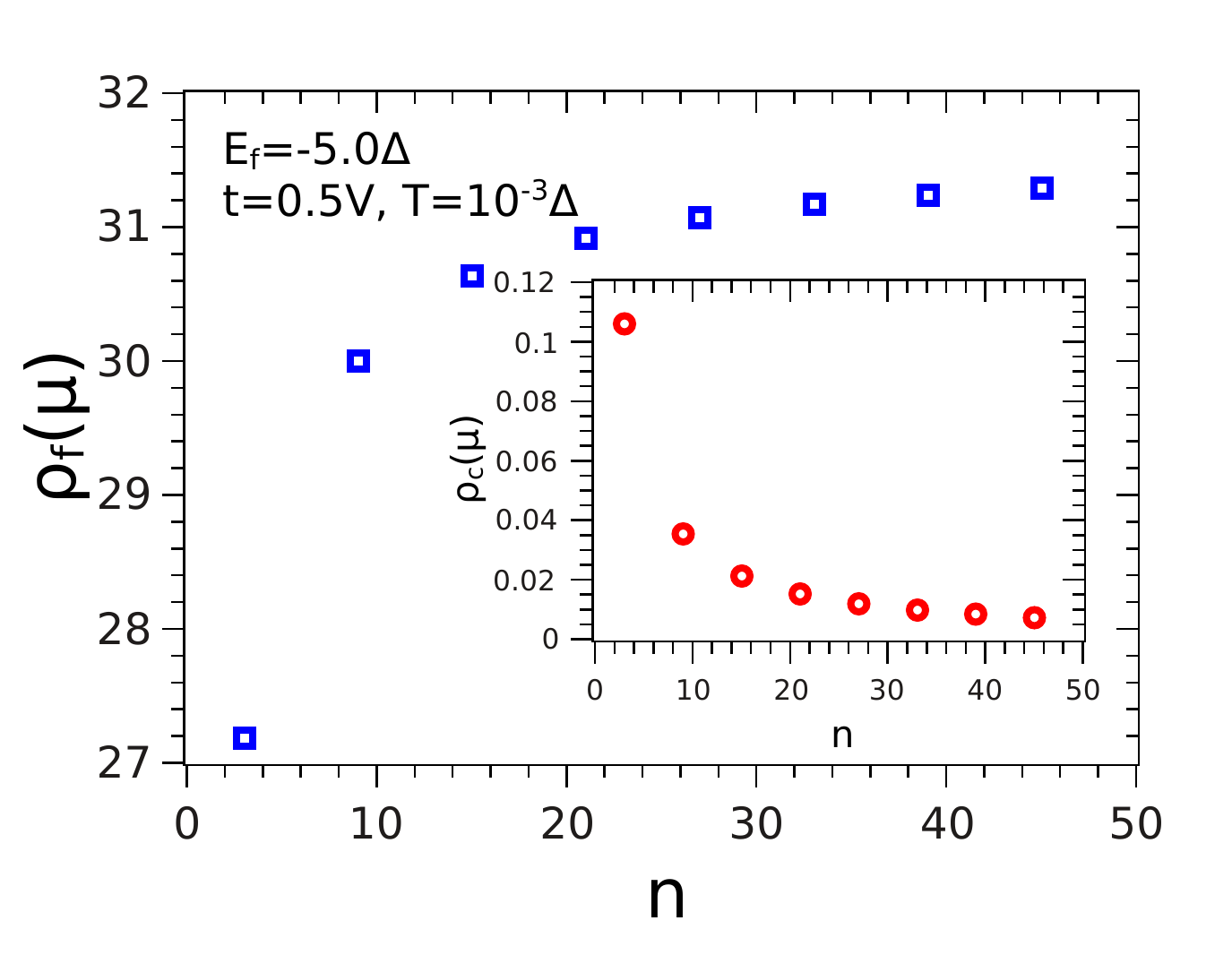}
	\caption{The density of states of the localized $f$ level $\rho_{f}$ at the chemical potential $\mu=0$ as a function of $n$, for $E_{f}=-5.0\Delta$, $\textbf{t}=0.5V$ and $T=0.001\Delta$.}
	\label{fig4}
\end{figure}

In Fig.(\ref{fig4}) we show the density of states of the localized electrons and of the conduction electrons at the chemical potential $\mu=0$. We can verify the inverse behavior between the height of the conduction band and the height of the Kondo peak.

\begin{figure}[htbp]
	\centering
	\includegraphics[width=7.5cm,height=6.0cm,angle=0.0]{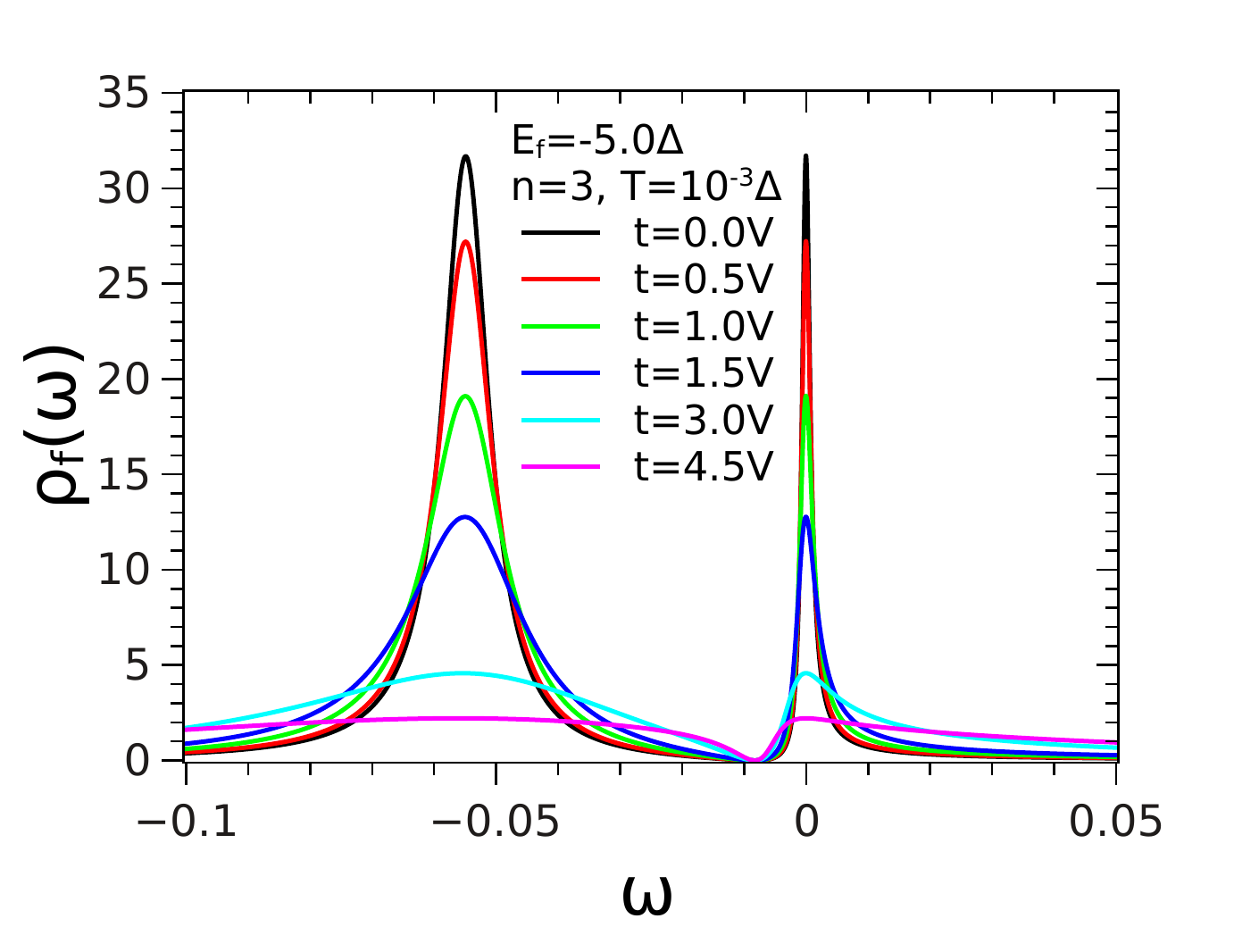}
	\caption{The density of states of the localized
		f level $\rho_{f}$ as a function of the $\omega$ (multiplied by $\Delta=0.01$) and different values of $\textbf{t}$. $E_{f}=-5.0\Delta$, $T=0.001\Delta$ and $n=3$.} 
	\label{fig5}
\end{figure}

In Fig.(\ref{fig5}) we present the density of states of the f-electrons localized at the QD for different values of $\textbf{t}$, which is the hopping between the nanotube and the QD, keeping all other parameters constant and for $n = 3$. 

We observed that an increase in $\textbf{t}$ produces a decrease in the height of the two peaks, which correspond to the localized level $Ef = -5.0\Delta$ (left peak) and the Kondo peak (right peak), localized at the chemical potential $\mu = 0$. As the interaction between the nanotube and the QD increases, the Kondo effect will be destroyed. This can be explained by the coupling geometry (T-shape ligation), which produces a destructive interference between the electrons of the nanotube and the QD.

The changes produced at the Kondo peak will produce changes in the transport properties of the system. These changes can be seen in the calculation of the conductance at site $0$ of the nanotube.

At low temperatures and bias voltage the electronic transport is coherent and the conductance is calculated according to the Landauer equation \cite{Kang2001}:

\begin{equation}
G=\frac{2e^{2}}{\hbar}\int\left(-\frac{\partial \Theta_{F}}{\partial\omega}\right)  S(\omega)d\omega,
\label{Eq18}
\end{equation}%
$\Theta_{F}$ is the Fermi function and $S(\omega)$ is the transmission probability of an electron with energy $\hbar\omega$ and is given by

\begin{equation}
S(\omega)=\gamma^{2}\mid G_{00}^{\sigma}\mid^{2},
\label{Eq19}
\end{equation}%
where $\gamma$ corresponds to the coupling strength, which is proportional to the kinetic energy of the electrons in the site $0$ and $G_{00}$ was obtained previously and is given by Eq. (\ref{Eq17}).

\begin{figure}[h!]
	\centering
	\includegraphics[width=7.5cm,height=6.0cm,angle=0.0]{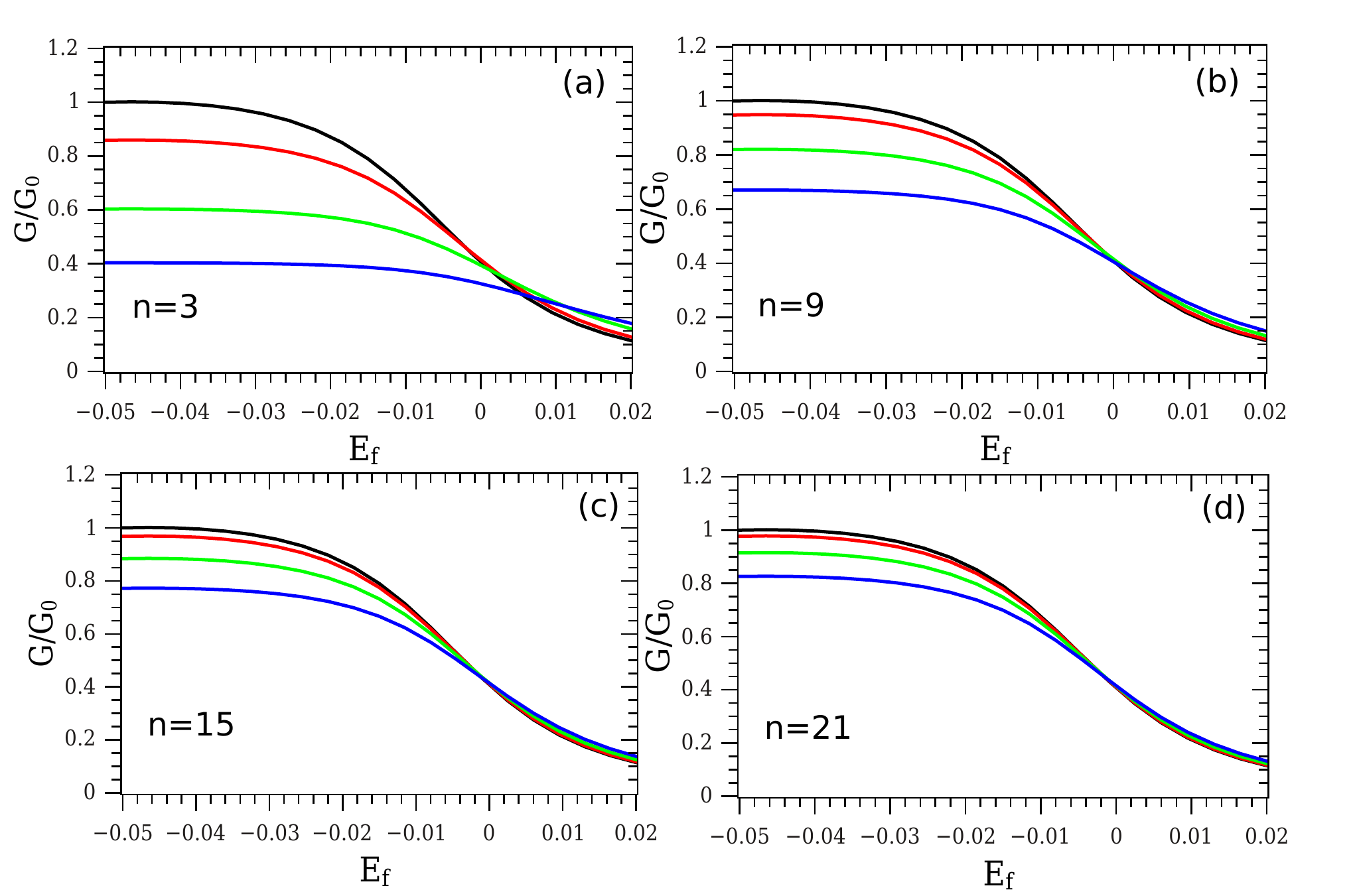}
	\caption{Condutance as a funtion $E_{f}$ (multiplied by $\Delta=0.01$) for different values of hopping $\textbf{t}=0.0V$ (black line); $\textbf{t}=0.5V$ (red line); $\textbf{t}=1.0V$ (green line) and $\textbf{t}=1.5V$ (blue line), with $T=0.001\Delta$ and for different values of $n$.} 
	\label{fig6}
\end{figure}

In the set of curves shown in Fig.(\ref{fig6}), we show the conductance calculated at site $0$ as a function of the level localized at the quantum dot $E_{f}$, for various values of $\textbf{t}$, which is the hopping that connects the QD to the carbon nanotube. We present four graphs, referring to the four values of $n=3,9,15,21$, which is proportional to the diameter of the nanotube.

First, we can observe in the four figures, for t = 0 (black line), the conductance has a unit value for $E_{f}=-5.0\Delta$, indicating that the Kondo resonance produces a ballistic transmission channel for the electrons coming from the leads. With the increase in $E_{f}$, the conductance drops smoothly until it approaches zero in the empty region ($E_{f}>0.0\Delta$). This behavior is known and is in agreement with results presented in previous studies \cite{aligia2002}.

As we connect the carbon nanotube to the (QD $+$ leads) system, that is, for $\textbf{t} = 0.5V, 1.0V, 1.5V$, we observe a destructive interference between the electrons coming from the metallic tube and the quantum dot. The Kondo peak is reduced in height (see Fig.(\ref{fig5})), implying a decrease in conductance in the Kondo region ($E_{f} <0.0\Delta$). This reduction in conductance is more important in tubes of smaller diameter ($n$ small). This can be explained by the height of the conduction band at the chemical potential $\mu$, which is higher in small tubes and smaller in larger tubes, as seen in Fig.(\ref{fig2}).

As the value of $n$ increases, that is, for increasingly larger diameters, the colored lines get closer and closer, indicating a reduction in destructive interference between the electrons from the nanotube and the QD, caused by the reduction in the amount of electrons coming from the nanotube. 
\begin{figure}[htbp]
	\centering
	\includegraphics[width=7.5cm,height=6.0cm,angle=0.0]{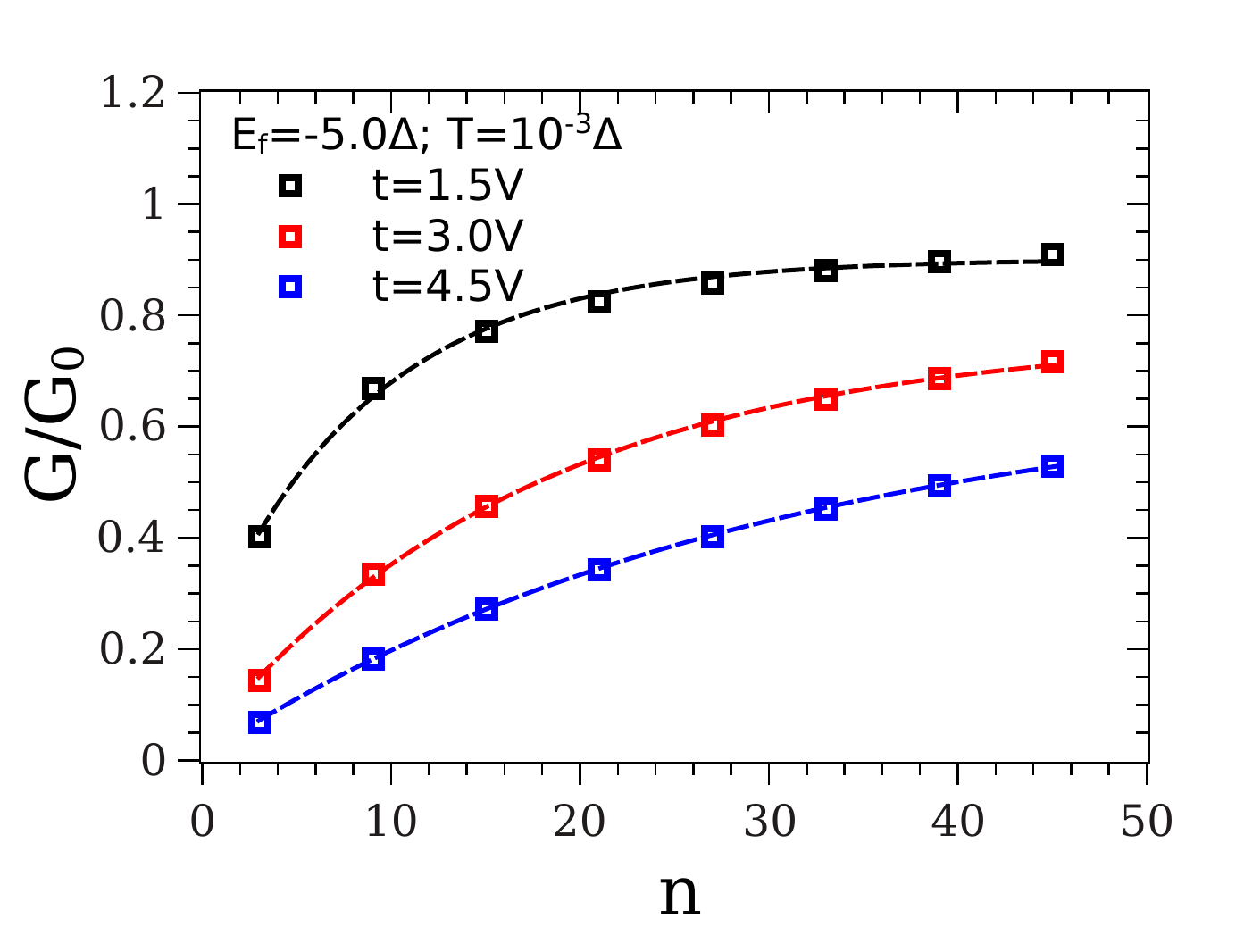}
	\caption{Condutance as a funtion $n$ for different values $\textbf{t}$. $E_{f}=-5.0\Delta$ and $T=0.001\Delta$.} 
	\label{fig10}
\end{figure}

In Fig.(\ref{fig10}) we represent the condutance as a function of $n$ for different values of hopping $\textbf{t}$. Although we do not present it in the figure, we know that for a zero hopping, that is, when the nanotube is disconnected from the system, the conductance has its unit value for any value of $n$. We can verify the influence of hopping in the destruction of the Kondo Effect. As the value of $\textbf{t}$ increases, the conductance curve shifts downwards.

\section{Conclusions}

In this study, we present a quantum dot system embedded in a ballistic channel and laterally coupled to a carbon nanotube of the zigzag type. The method used was the atomic approximation, which has great reliability in the low temperature regime, satisfying Friedel's sum rule.

We present density of states curves, both for the conduction band of the nanotube and for the electrons localized $f$. We found that an increase in the diameter of the nanotube produces a decrease in the height of the conduction band in the chemical potential $\mu=0$. This decrease, in turn, produces an increase in the height of the Kondo peak, located in the same energy region ($\mu=0$). This inverse relationship is due to the geometry of the connection between the nanotube and the quantum dot, which are connected laterally.

We also looked at the role of hopping $\textbf{t}$ that links the quantum dot to the carbon nanotube. We verified through the density of electron states $f$ that the Kondo Peak decreases with the increase of $\textbf{t}$, again indicating the destructive interference characteristic between the electrons from the nanotube to the quantum dot, when the coupling is lateral.

As the variation of $n$ and $\textbf{t}$ changes the height of the Kondo peak, that is located in the chemical potential, the transport properties will also be affected. We observed this effect on the conductance curves.

We found that when $\textbf{t} = 0$, that is, when the nanotube is disconnected from the system, the conductance is ballistic at $E_{f} = -5.0\Delta$, indicating that the Kondo effect is present and that the Kondo peak produces a conduction channel for the electrons coming of leads. As $\textbf{t}$ increases, the conductance tends to decrease, as the electrons coming from the nanotube conduction band become more and more abundant. We also found that small diameter nanotubes are more sensitive to changes in hopping for conductance calculation.

\vspace{1.0cm}
\textbf{ACKNOWLEDGEMENTS}
We would like to express our gratitude to Professor M. S. Figueira for his encouragement and helpful consultation along the development of this study
This work was partially supported by CNPq - Grants: 407462/2018-0 and 306569/2018-3 
(Brazilian Research Agency).

\end{document}